\documentstyle[12pt]{article}
\begin{document}
\title{ Moment Density of Zipoy's Dipole Solution}
\author{L. Fern\'andez-Jambrina\\
Departamento de F\'{\i}sica Te\'orica II,
\\Facultad de
Ciencias F\'{\i}sicas,
\\Universidad Complutense 
\\28040-Madrid, Spain}
 \date{}
\maketitle 
\abstract{Zipoy used spheroidal coordinates to construct a family of static
axisymmetric gravitational fields included in Weyl's class. We calculate the
mass moment source for the dipole solution using techniques previously
developed for magnetic and angular momentum densities and compare it with the
Newtonian analog studied by Bonnor and Sackfield.}    
\vspace{0.5cm}

\hfil PACS: 04.20.Cv, 04,20.Jb \hfil
\newpage 

\section{Introduction}

In \cite{Bon} Bonnor and Sackfield applied potential theory to analyse some
metrics of the family obtained by Zipoy \cite{Zip}. They calculated the mass
density for the monopole solution from the discontinuities of the derivatives
of the Weyl potential \cite{Weyl}, taken as a Newtonian potential, and also
they calculated the moment density for the dipole solution
from the discontinuity of the potential itself, considering these jumps
as surface sources for the gravitational fields. While their result for the
monopole solution was also grounded on the thin surface layer formalism
presented by Israel in \cite{ISR} and therefore it had an interpretation
within the framework of General Relativity, their analysis of the dipole
solution was carried out using only a Newtonian flat spacetime analog,
concluding that a sheet of dipoles was the only source present. In this paper
a justification for that result will be given using a modification suited for
mass dipoles of the formalism derived previously in \cite{first} and
\cite{second} for rotation and electromagnetic moments in General Relativity.

Section 2 is devoted to the Weyl static axisymmetric family of vacuum
solutions of the Einstein equations using in its derivation the differential
form approach applied in \cite{ch} and \cite{tes} to vacuum fields. Zipoy's
spheroidal metrics are dealt with in Section 3. In Section 4 use is made of
the formalism introduced previously to provide a way of calculating dipole
densities out of the discontinuities of the metric functions. This method is
applied to the Zipoy dipole solution in Section 5. A brief discussion of the
results is provided in Section 6.

\section{The Weyl Static Family}

The Weyl static axisymmetric vacuum solutions were discovered soon after the
publication of the General Theory of Relativity \cite {Weyl}. It is the
general solution for metrics with this symmetry and allows the construction
of metrics from solutions of a flat spacetime Laplace equation.

We shall introduce the relevant equations
for the problem in the form of a exterior system. Our start point is an
orthonormal vierbein $\{\theta^0,\theta^1,\theta^2,\theta^3\}$ such that the
metric takes this form:

\begin{equation}
^4g=-\theta^0\otimes\theta^0+\theta^1\otimes\theta^1+\theta^2\otimes\theta^2
+\theta^3\otimes\theta^3
\end{equation}

The forms $\theta^0$ and $\theta^1$ lie on the space spanned by the orbits of
the two Killing fields, $\partial_t$ and $\partial_{\phi}$, while the other
two forms, $\theta^2$ and $\theta^3$, lie on the orthogonal space.

The equations stating that the frame is torsion-free
(\ref{eq:uno}-\ref{eq:cuatro}), together with its integrability conditions
(\ref{eq:cWeyl}-\ref{eq:acc}) and Einstein's equations
(\ref{eq:Weyl}-\ref{eq:Ray}) can be shown to be \cite{ch}, \cite{tes}:

\begin{equation} 
d \theta^0 = a \wedge \theta^0 \label{eq:uno}
\end{equation} 
\begin{equation}
 d \theta ^1 = (b - a) \wedge \theta ^1 
\end{equation}
\begin{equation}
d\theta^2=-\nu\wedge\theta^3 
\end{equation}
\begin{equation}
d\theta^3=\nu\wedge\theta^2 \label{eq:cuatro}
\end{equation}
 
\begin{equation} 
d b  =  0 \label{eq:cWeyl}
\end{equation} 

\begin{equation} 
d a = 0 \label{eq:acc}
\end{equation}

\begin{equation}
d * b + b \wedge * b = 0\label{eq:Weyl}
\end{equation}

\begin{equation}
d * a + b \wedge * a =0\label{eq:Ray} 
\end{equation}
 The symbol $*$ stands for the Hodge duality operator in the $\theta^2-
\theta^3$ space, that is $*\theta^2=\theta^3$ and $*\theta^3=-\theta^2$. The
equations governing the connection $\nu$ in the $\theta^2-\theta^3$ space have
not been written since they can be integrated by quadratures after the previous
exterior system has been solved and are not of interest for our purposes here. 
 
The equations \ref{eq:uno}-\ref{eq:Weyl} can be formally integrated to yield
the metric in canonical coordinates $\{t,\phi,z,\rho\}$ and therefore the
meaning of the one-form $b$ as the differential of the logarithm of
the volume of the $\theta^0-\theta^1$ space is clear:

\begin{equation}
a=dU\ \ \ \ \ b=d\ln\rho\ \ \ \ \ *b=-\rho^{-1}dz\label{eq:muchas}
\end{equation}

\begin{equation}
^4g=-e^{2U}dt^2+e^{-2U}\{\rho^2d\phi^2+e^{2k}(d\rho^2+dz^2)\}\label{eq:metric}
\end{equation}

The remaining equation, \ref{eq:Ray}, is analagous to the Raychaudhuri
equation for the accelerarion of a fluid. Therefore $a$ can be viewed as an
acceleration one-form. After substitution of  \ref{eq:muchas} it takes the
form of a reduced three dimensional Laplace equation:

\begin{equation}
d*dU+\rho^{-1}d\rho\wedge *dU=0
\end{equation}

Or in explicit Weyl canonical coordinates:

\begin{equation}
U_{\rho\rho}+U_{zz}+\frac{1}{\rho}U_\rho=0
\end{equation}

As we had already stated, Newtonian potential functions, $U$, yield after
substitution in the metric \ref{eq:metric} exact solutions of Einstein's
equations. But the analogy finishes here and no further interpretation can be
brought from this fact. As an example, the classical potential for a finite
rod with constant density produces the Schwarzschild metric, the relativistic
monopole solution. 

If we want the metric to be asymptotically flat, then $U$ should be at most
monopolar at infinity, that is:

\begin{equation}
U=-\frac{m}{r}+O\left(\frac{1}{r^2}\right)
\end{equation}
We shall restrict to these metrics from now on.

\section{Zipoy's Metrics}

Zipoy's metrics belong to Weyl class \cite{Zip} and are constructed from
solutions of the Laplace equation in  oblate spheroidal
coordinates $\{r,\theta\}$ instead of using pseudocilindrical coordinates
$\{\rho,z\}$:

\begin{equation}
z=r\cos\theta\ \ \ \ \ \rho=\sqrt{r^2+a^2}\sin\theta
\end{equation}
where $a$ is a constant and the new coordinates range as usual: 

\begin{equation}
0\leq r<\infty\ \ \ \ \ \ \ \ \ 0\leq\theta\leq\pi
\end{equation}

Note that events with coordinates $(t,\phi,r=0,\theta)$ and $(t,\phi,r=0,
\pi-\theta)$ have the same canonical coordinates. Therefore we shall identify
them instead of attempting more complicated interpretations.

In Bonnor and Sackfield's paper, $\theta$ is a latitude angle instead of a
colatitude, as it is taken in this paper, and also we make use of the radius
$r$ instead of the coordinate $u$ related to the previous ones by $r=a\sinh
u$. We have done this since their interpretation as asymptotic spherical
coordinates is simpler.

The reduced Laplace equation in these coordinates takes the following form:

\begin{equation}
(r^2+a^2)U_{rr}+2rU_r  +U_{\theta\theta}+\cot\theta U_\theta=0
\end{equation}

We can construct solutions for this equation of the form
$U_n=R_n(r)P_n(\cos\theta)$ where $P_n$ stands for the Legendre polynomials.
The index $n$ runs from zero to infinity allowing us to classify the
Newtonian potentials $U$ and the metrics derived from them using a multipole
expansion. We shall be concerned with the solution for $n=1$, Zipoy's dipole
solution.

\section{Dipole Sources}

Since the aim of this paper is just the interpretation of Zipoy's dipole
solution, we shall restrict ourselves to  asymptotically dipolar static
axisymmetric vacuum solutions of the Einstein equations. For this we mean
that the metric function $U$ must be:

\begin{equation}
U=\frac{M\cos\theta}{r^2}+O\left(\frac{1}{r^3}\right)\label{asU}
\end{equation}
where $M$ is the total mass dipole.

Now, we shall proceed as in \cite{first} and \cite{second} and consider the
two formal expressions for the acceleration form arising from the integration
of equations \ref{eq:acc} and \ref{eq:Ray}:

\begin{equation}
a=dU=-\rho^{-1}*d\Lambda
\end{equation}

The asymptotic behaviour of $U$ imposes a condition at infinity for the newly
introduced function $\Lambda$:

\begin{equation}
\Lambda=-\frac{M\sin^2\theta}{r}+O\left(\frac{1}{r^2}\right)\label{asL}
\end{equation}

In order to have a dipole source for the gravitational field, it is assumed
that the function $U$ is discontinuous across a closed surface $S$. This
condition on $S$ means no restriction, since the difference between the
values of $U$ on either side of the surface can be null on some parts of it.

Taking into account both expressions for the acceleration, we integrate the
scalar product of $a$ and the differential of the Weyl coordinate $z$ with a
factor $e^U$. The domain of integration will be the space $V_3$ orthogonal to
$\theta^0$, whose metric is $g=^4g+\theta^0\otimes\theta^0$. The square root
of the determinant of this metric is $\sqrt{g}=\rho e^{2k-3U}$.

\begin{eqnarray}
0&=&\int_{V_3}\sqrt{g}e^U<[a+*(*a)],dz>dx^1dx^2dx^3=\nonumber\\&=&\int_{V_3}
\sqrt{g}e^U<[dU+\rho^{-1}*d\Lambda],dz>dx^1dx^2dx^3\label{eq:plasta}
\end{eqnarray}

The Weyl coordinate $z$ satisfies the following differential equation arising
from \ref{eq:Weyl}, since $\sqrt{^4g}=e^U\sqrt{g}$:

\begin{equation}
\partial_\mu\{\sqrt{g}e^Ug^{\mu\nu}\partial_\nu z\}=0
\end{equation}

Therefore the integrand in \ref{eq:plasta} can be shown to be a total
derivative and the expression can be reduced to a surface integral. 

\begin{eqnarray}
0=\int_{V_3}\partial_\mu\{\sqrt{g}Ue^Ug^{\mu\nu}\partial_
\nu z+[\mu\nu]\Lambda\partial_\nu z\}dx^1dx^2dx^3
\end{eqnarray}

Actually,
since $U$ is discontinuous, the integral has to be split into two pieces, 
$V_3^+$ and $V_3^-$, respectively the outer and inner sections of $V_3$
referred to the surface $S$. The boundary of $V_3^-$ is formed by the surface
$S$ and the boundary of $V_3^+$ consists of the same surface and the 2-sphere
at infinity $S^2(\infty)$. The integral over $S^2(\infty)$ can be
straightforwardly carried out, taking into account the asymptotic conditions
\ref{asU} and \ref{asL} and that the metric is assumed to be asymptotically
flat:

Hence the integral defining the total mass dipole is just:
\begin{equation}
4\pi M=\int_{S^+}dS^+\{Ue^Ug^{\mu\nu}n_\mu\partial_
\nu z\}^+-\int_{S^-}dS^-\{Ue^Ug^{\mu\nu}n_\mu\partial_ 
\nu z\}^-
\end{equation}

Due to the fact that the metric is discontinuous -the potential $U$ is also
part of the metric and of the unitary normal, $n$, to the surface $S$-, it
seems in principle not possible to reduce the previous equation to one
integral of the jump of $U$. But it can be shown, for instance, that, if the
discontinuity takes places on a disk on the plane $z=0$, metric factors like
$e^U$ cancel each other and therefore the differential element of mass dipole
is perfectly defined. 

\section{The Dipole Solution's Source}

The aim of this paper was the interpretation of Zipoy's dipole metric with
the formalism developed in the previous section. As it has already been said,
this metric arises from the substitution of the dipolar solution of the
Laplace equation in oblate spheroidal coordinates in the Weyl metric. This
solution can be shown to be:

\begin{equation}
U=\gamma\{1-\frac{r}{a}\arctan(\frac{a}{r})\}\cos\theta
\end{equation}
where $\gamma$ is a constant.

It is easily seen that this function behaves for large values of $r$ like a
dipolar potential whose moment is $M=\frac{1}{3}\gamma a^2$.

\begin{equation}
U=\frac{1}{3}\gamma a^2\frac{\cos\theta}{r^2}+O\left(\frac{1}{r^4}\right)
\end{equation}

 The remaining
metric function, although it will not be needed is:

\begin{equation}
e^{2k}=\left(\frac{r^2+a^2\cos^2\theta}{r^2+a^2}\right)^{-\gamma^2}\exp\{-U^2
\tan^2\theta-\gamma^2\sin^2\theta[\arctan(\frac{a}{r})]^2\}
\end{equation}

As we have identified on the surface of null $r$ coordinate points whose
colatitude is $\theta$ with those that have $\pi-\theta$, it happens that,
when we reach that surface from the subspace $z\geq 0$, the potential takes
the value $U=\gamma\cos\theta$, but if we arrive with $z\leq 0$, then it is
equal to  $U=-\gamma\cos\theta$, $\theta$ ranging from $0$ to $\frac{\pi}{2}$
after the identification of events \cite{Bon}.

Therefore the potential meets a jump at the disk $z=0$, $\rho\leq a$ ($r=0$,
$\theta\leq\frac{\pi}{2}$):

\begin{equation}
[U]=2\gamma\cos\theta=2\gamma\sqrt{1-\frac{\rho^2}{a^2}}
\end{equation}

The disk $z=0$ has the following surface element and normal:

\begin{equation}
dS=\rho e^{k-2U}d\rho d\phi\ \ \ \ \ \ n=e^{U-k}\partial_z
\end {equation}

Therefore the product $e^Un^\mu\partial_\mu z dS$ does not depend on $U$ and
takes the same value on both sides of the disk $S$, as we anticipated in the
previous section.

Hence the integral defining the total mass dipole can be written as:

\begin{equation}
M=\frac{1}{4\pi}\int_S dS\
[U]\ e^Un^\mu\partial_\mu z = \frac{1}{4\pi}\int^{2\pi}_0d\phi\int^a_0d\rho\
\rho\ 2\gamma\sqrt{1 -\frac{\rho^2}{a^2}}=\frac{1}{3}\gamma a^2\label{dip} 
\end{equation}

The mass dipole takes the value expected from the asymptotic form of the
metric and the differential element of the mass moment is then:

\begin{equation}
dM=\frac{1}{4\pi}\ \rho\ 2\gamma\sqrt{1
-\frac{\rho^2}{a^2}}d\phi d\rho\label{dden}
\end{equation}
in full accordance with Bonnor and Sackfield's Newtonian result.

\section{Discussion}

In the previous section the differential surface element
for the mass moment source of Zipoy's dipole metric 
 is constructed from the discontinuity of
the metric function $U$. This result coincides with the one
obtained by Bonnor and Sackfield \cite{Bon} using classical potential theory,
since the curved spacetime elements in the integral \ref{dip}
arising from the metric cancel each other. Therefore the full relativistic
theory and its classical Newtonian aproximation yield the same output and
Bonnor and Sackfield's result is confirmed within the framework of General
Relativity. 

Something similar happens with the angular momentum density for Kerr's
metric \cite{first}, where oblate spheroidal coordinates are also employed.
As it is stated in \cite{Bon}, the potential $U$ is due only to the sheet of
dipoles described in \ref{dden}, that is, there is no contribution of higher
order multipole sheets to the gravitational field and therefore the source is
fully described, at least in the Newtonian limit. Again this is the case for
Kerr's metric. 

It remains to be seen if this also happens when the full
relativistic theory is taken into account.

\noindent
{\it The present work has been supported in part by DGICYT Project PB92-0183;
L.F.J. is supported by a FPI Predoctoral Scholarship from Ministerio de
Educaci\'{o}n y Ciencia (Spain). The author wishes to thank F.J. Chinea and L.M.
Gonz\'alez-Romero for valuable discussions.}

\end{document}